# Feedback Induced Death in Coupled Oscillators

Ming Luo

School of Aerospace Engineering and Applied Mechanics, Tongji University, Shanghai, China 200092

Abstract. We investigate oscillation death in systems of coupled nonlinear oscillators with feedback loop. We find that feedback results in oscillation death both in small sets or large ensembles. More importantly, the death zone in parameter space is significantly enlarged and oscillation death could occur even in coupled identical oscillators in the presence of feedback. We find that there are two different ways to oscillation death, namely desynchronization and completely synchronization induced oscillation death. Feedback induced oscillation death may be used to suppress unexpected oscillations, e.g., in chaotic laser arrays.

Systems of interacting nonlinear oscillators have been extensively studied because both of their abundance and of their significance in many areas of science and technology [1]. Based on the nature of individual units, couplings and network topology, such systems can exhibit variety of emergent phenomena. One of the most important collective behaviors is synchronization, which has been widely studied in many different contexts such as chemical reaction systems [2], laser systems [3], ecosystems [4], biological systems [5], etc. Another intriguing collective phenomenon is oscillation death, which has been widely reported to exist in chemical, physical, and biological systems etc. [6, 7]. It is relevant to certain physiologies and pathologies in biological oscillator networks [8, 9]. Early studies showed that oscillation death could only occur in systems with strong enough coupling and large enough parameter mismatch [7]. The work of Reddy et al. demonstrated that oscillation death occurs even in identical systems with time-delayed coupling[10]. Since then, considerable interests have been expressed to the amazing phenomenon. Up to now, several mechanisms, such as nonlinear coupling, the coupling through dissimilar or conjugate variables, etc., have been proposed for the occurrence of oscillation death in identical systems even without time delay [11].

On the other hand, previously studies have shown that feedback plays a crucial role in collective behaviors. As an integral component, inherent feedback has been reported to have considerable effects on collective dynamic behaviors. For example, corticothalamic feedback has been shown to enhance synchronization of the firing of lateral geniculate nucleus (LGN) cells [12]. Circadian rhythms are controlled by the interactions within the circadian clock in the brain and feedback from other brain parts and locomotive activities [13]. External feedback has also been widely used to control the behavior of complex rhythms [14]. For instance, global feedback is adopted to control the coherence of spatiotemporal dynamics in Belousov-Zhabotinsky (BZ) systems [15]. Well-designed feedback signals are also applied to destroy unwanted synchronization, with promising applications in medical science to suppress pathological neural synchrony in several neurological diseases, such as Parkinson's diseases and epilepsy [14, 16]. However, little attention is given to oscillation death due to internal or external feedback till now.

Here we study feedback induced oscillation death in coupled oscillators. The effects of feedback on the death zone of two coupled oscillators are examine. With increase of feedback gain, the death zone changes, extending not only to zero parameter mismatch, but also to zero coupling. Then, large ensembles of coupled oscillators is considered. We found that once feedback gain is large then some threshold, oscillation death occurs with strong enough coupling strength. However, the death zone is independent of feedback gain, i.e., it does not change with feedback.

We analysis a pair of coupled limit-cycle oscillators with feedback [16, 17]

$$\dot{z}_1(t) = (\alpha + i\omega_1 - |z_1|^2)z_1 + \frac{\varepsilon}{2}(z_2(t) - z_1(t)) + u(t)$$
$$\dot{z}_2(t) = (\alpha + i\omega_2 - |z_2|^2)z_2 + \frac{\varepsilon}{2}(z_1(t) - z_2(t)) + u(t)$$
(1)

where $z_{1,2}$ are complex numbers which represent the states of the two oscillators at time $t$, $\omega_{1,2}$ are the natural frequencies, $\alpha > 0$ determines the natural individual radii, $\varepsilon \geq 0$ is the coupling strength, $u(t) = -gZ(t)$ with $Z(t) = (z_1 + z_2)/2$ representing average behavior of two oscillators, and $g$ is feedback gain [16].

It is clear that the origin $z_j = 0, j = 1, 2$ is always a fixed point of the closed loop system (1). The characteristic equation for the fixed point is

$$\lambda^2 - [2\alpha - \varepsilon - g + i(\omega_1 + \omega_2)]\lambda + \frac{(2\alpha - \varepsilon - g)^2}{4} - \omega_1\omega_2 - \frac{(\varepsilon - g)^2}{4} + \frac{2\alpha - (\varepsilon + g)}{2}(\omega_1 + \omega_2)i = 0 \quad (2)$$

where $\lambda$ is the complex eigenvalue. The marginal stability curves or the critical curves are thus obtained by setting $\lambda = i\beta$

$$\beta(2\alpha - \varepsilon - g) - \frac{2\alpha - \varepsilon - g}{2}(\omega_1 + \omega_2) = 0 \tag{3}$$

$$\beta^2 - \beta(\omega_1 + \omega_2) - \frac{(2\alpha - \varepsilon - g)^2}{4} + \omega_1\omega_2 + \frac{(\varepsilon - g)^2}{4} = 0 \tag{4}$$

By (3), we get $2\alpha - (\varepsilon + g) = 0$, and $\beta = (\omega_1 + \omega_2)/2$. Substituting for $\beta$ in (3), we obtain $\Delta^2 + 4\alpha^2 - 4\alpha\varepsilon - 4\alpha g + 4\varepsilon g = 0$, where $\Delta = |\omega_1 - \omega_2|$. The critical curves are

$$2\alpha - (\varepsilon + g) = 0, \text{ and } \Delta^2 - 4\alpha\varepsilon + 4\varepsilon g - 4\alpha g + 4\alpha^2 = 0 \tag{5}$$

Without feedback, i.e. $g = 0$, the critical curves are

$$\varepsilon = 2\alpha, \text{ and } \varepsilon = \alpha + \Delta^2/4\alpha \tag{6}$$

The death region in $(\varepsilon, \Delta)$-space without feedback is shown in Fig. 1a (yellow region) [18]. We can see that oscillation death can only occurs with both strong enough coupling and wide enough frequency mismatch [18]. Moreover, there is a "death valley", namely for certain frequency mismatch, oscillation death occurs only with intermediate coupling ($2\alpha < \varepsilon < \alpha + \Delta^2/4\alpha$). When the coupling increases further and $\varepsilon > \alpha + \Delta^2/4\alpha$, the two oscillations synchronize again.

Evolution of death region in $(\varepsilon, \Delta)$-space with the feedback gain $g$ is show in Fig. 1. For small feedback gain $g < \alpha$, the critical curves are the line $\varepsilon = 2\alpha - g > 0$ and the right-open parabola $\varepsilon = \frac{\Delta^2}{4(\alpha - g)} + \alpha$. The death region is similar to that of the uncontrolled case. For $g = \alpha$, the death zone is determined by two lines $\varepsilon = \alpha$, and $\Delta = 0$. As shown in Fig. 1b (yellow region), oscillation death occurs for any frequency mismatch as long as the coupling strength $\varepsilon > \alpha$. For $g > \alpha$, the parabola (5) opens to the left. When $\alpha < g < 2\alpha$, the critical curves are the line $\varepsilon = 2\alpha - g > 0$ and the left-open parabola $\varepsilon = \frac{\Delta^2}{4(\alpha - g)} + \alpha$ (Fig. 1c, yellow region). When $g \geq 2\alpha$, the critical curves are the line $\varepsilon = 0$ and the left-open parabola $\varepsilon = \frac{\Delta^2}{4(\alpha - g)} + \alpha$ (Fig. 1d, yellow region). For the intersection of the critical curves with $\Delta$-axis is $\Delta = 2\sqrt{4\alpha(g - \alpha)}$. Thus, the death zone shrinks to $\varepsilon > \alpha$ with increasing feedback gain (Fig. 1d).

The next question is what happens in the case of large ensemble of oscillators, which is of great scientific and practical significance. Thus, we consider the following ensemble of all-to-all coupled limit-cycle oscillators with feedback [16, 17]

$$\dot{z}_j = (\alpha + i\omega_j - |z_j|^2)z_j + \frac{\varepsilon}{N}\sum_{k=1}^{N}(z_k - z_j) + u(t) \qquad (7)$$

where $j = 1, \cdots, N$ and again the feedback is simply taken as $u(t) = -gZ(t)$ with $Z(t) = \frac{1}{N}\sum_{k=1}^{N}z_k$ [16]. For $g = 0$ (without feedback), equation (7) reduces to the model which have been extensively studied [19, 20]. Similar to the case of $N = 2$, death can only occur for sufficiently large frequency mismatch and sufficiently strong coupling. Furthermore, the death zone is dependent on the distribution of individual frequencies. In Fig. 2a, we reproduce the death region of Mirollo and Strogatz with uniformly distributed natural frequencies within $[-\gamma, \gamma]$. The boundary is defined by $\gamma \cot(\gamma/\varepsilon) + \alpha - \varepsilon = 0$ and $\varepsilon > \alpha$ [19].

A simple way is used to figure out the death zone of the closed loop system (7). In the case of large ensembles, $N \to \infty, \frac{z_j}{N} \to 0, \frac{1}{N}\sum_{k=1}^{N}z_k \approx \frac{1}{N}\sum_{k=1, k \neq j}^{N}z_k$. Equation (7) can be approximated by the following [17]

$$\dot{z}_j = (\alpha - \varepsilon + i\omega_j - |z_j|^2)z_j + \frac{\varepsilon}{N}\sum_{k=1, k \neq j}^{N}z_k - \frac{g}{N}\sum_{k=1, k \neq j}^{N}z_k \qquad (8)$$

From the control point of view, the negative feedback $u(t) = -gZ(t)$ stabilizes the mean field $Z(t)$. With large enough feedback gain $g$, $Z(t) \to 0$, and then $u(t) \to 0$. Equation (8) can be approximated [17]

$$\dot{z}_j = ((\alpha - \varepsilon) + i\omega_j)z_j - |z_j|^2 z_j \qquad (9)$$

From equation (9), it is clear that both the collective behaviors and dynamics of individual oscillators are determined by parameters $\alpha$ and $\varepsilon$. Let $z_j = r_j e^{i\theta_j}$, equation (9) becomes

$$\dot{\theta}_j = \omega_j \quad \text{and} \quad \dot{r}_j = r_j(\alpha - \varepsilon - r_j^2) \qquad (10)$$

For $\alpha > \varepsilon$, equation (10) has unstable equilibrium points zero $r_j = 0$ and stable fixed points $r_j = \sqrt{\alpha - \varepsilon}$. All the individual units oscillate with their natural frequencies $\omega_j$ and radii $\sqrt{\alpha - \varepsilon}$, i.e. the ensemble are desynchronized. For $\alpha \leq \varepsilon$, equation (10) has only one stable equilibrium point zero. All the individual oscillations are quenched. The ensemble enter into the oscillation death zone. Fig. 2b shows bifurcation diagrams in $(\gamma, \varepsilon)$-space of equation (7) with strong feedback. Compared to Fig. 2a, the death zone extends significantly. As long as $\varepsilon > \alpha$, oscillation death occurs even in coupled identical oscillators, which is in great contrast with the case without feedback. Compared to Fig. 1d, we can see that the death zone does not change with feedback gain

$g$. However, it never occurs for $\varepsilon < \alpha$.

Based on the discussion above, we come up with a general way to oscillation death in an ensemble of coupled oscillators, referred to as desynchronization induced oscillation death. There are two basic factors for this type of oscillation death to happen, namely desynchronizing mechanism such as time delay in the coupling, frequency mismatch, etc., which prevents the ensemble from synchronization [10, 16, 19], and stabilizing mechanism such as self-feedback, which quenches individual oscillations [19, 20, 21]. Fig. 3a shows an example of oscillation death induced by feedback in diffusively coupled identical oscillators. The main feature of this type of oscillation death is that the oscillators fall into the state of zero amplitude incoherently.

Feedback can result in another type of oscillation death in coupled identical oscillators, which is called completely synchronization induced oscillation death. An example is shown in Fig. 3b. Without feedback, the oscillators are full-synchronized (Fig. 3b, $t < 25$). When the feedback is set on, they all collapse into the same equilibrium states (Fig. 3b, $t > 25$). The main feature of this type of oscillation death is the ensemble are full-synchronized all the time. We note that this oscillation death occurs even for $\varepsilon < \alpha$, which however is unstable.

The above death mechanism is general and is applicable to coupled chaotic oscillators. Take an ensemble of coupled chaotic Rossler oscillators as an example [22],

$$\begin{aligned}\dot{x}_j &= -\omega_j y_j - z_j \\ \dot{y}_j &= \omega_j x_j + a y_j + \frac{\varepsilon}{N}\sum_{k=1}^{N}(y_k - y_j) \\ \dot{z}_j &= b + x_j z_j - c z_j\end{aligned} \quad (11)$$

where parameters $a = 0.2, b = 0.4, c = 8.5$. $\omega_j$ are randomly selected. Oscillators are all-to-all coupled via $y_j$ variables. For certain distribution of $\omega_j$, the ensemble (11) synchronize when the coupling strength $\varepsilon$ is large enough (Fig. 4a).

In this example, a washout filter aided feedback is applied. We assume that the mean activities $Y = \sum_{j=1}^{N} y_j$ is measured and thus the filter is of the following form [21]

$$\begin{aligned}\dot{w}(t) &= Y - d w(t) \\ v(t) &= Y - d w(t)\end{aligned} \quad (12)$$

where $w(t)$ and $v(t)$ are the state variable and output of the filter, respectively. $d$ is the filter constant. The feedback is $u(t) = -gv(t)$ and is administered to all the oscillators through variables $y_j$, $g$ is the feedback gain.

With large enough feedback gain $g$, the overall oscillations would be suppressed [17], and the mean field $(X,Y,Z) \to (X_0, Y_0, Z_0)$ [16], where $X = \sum_{j=1}^{N} x_j$, $Y = \sum_{j=1}^{N} y_j$ and $Z = \sum_{j=1}^{N} z_j$. The coupling $\frac{\varepsilon}{N} \sum_{k=1}^{N} (y_k - y_j) = \varepsilon Y - \varepsilon y_j \to \varepsilon Y_0 - \varepsilon y_j$. Furthermore, feedback signal vanishes by washing out the steady state $Y_0$ [21]. Thus, equation (11) is approximated by the following [17]

$$\begin{aligned} \dot{x}_j &= -\omega_j y_j - z_j \\ \dot{y}_j &= \omega_j x_j + (a-\varepsilon) y_j + \varepsilon Y_0 \\ \dot{z}_j &= b + x_j z_j - c z_j \end{aligned} \quad (13)$$

Each oscillator has two fixed points $F_\pm$ located at

$$(x_\pm, y_\pm, z_\pm) = \left( \frac{c\omega_j - \varepsilon Y_0 \pm \sqrt{c^2 \omega_j^2 - 4ab + 2c\varepsilon \omega_j Y_0 + \varepsilon^2 Y_0^2}}{2\omega_j}, -\frac{c\omega_j - \varepsilon Y_0 \pm \sqrt{c^2 \omega_j^2 - 4ab + 2c\varepsilon \omega_j Y_0 + \varepsilon^2 Y_0^2}}{2a\omega_j}, \frac{c\omega_j + \varepsilon Y_0 \pm \sqrt{c^2 \omega_j^2 - 4ab + 2c\varepsilon \omega_j Y_0 + \varepsilon^2 Y_0^2}}{2a} \right), \quad (14)$$

which are unstable saddle-nodes for $\varepsilon > a$. When $\varepsilon > a$, the fixed points $F_-$ become stable. All the oscillators are attracted to their stable equilibrium states. The ensemble undergo oscillation death. Bifurcation diagram of collective behaviors of the closed-loop system (11) and (12) in $(\gamma, \varepsilon)$-space is shown in Fig. 4b. Fig. 4c shows an example of an ensemble of coupled identical Rossler oscillators. Both the mean field (Fig. 4c red line) and the oscillators (Fig. 4c, blue line) converge to the same fixed point. Fig. 4e shows an example of an ensemble of coupled non-identical oscillators. The mean field and the oscillators converge to different fixed points. Counterintuitively, rhythmic activities of all the individual units are suppressed with a vanishing feedback [23].

In conclusion, feedback result in oscillation death both in small sets or large ensembles of coupled oscillators, even in systems of identical oscillators. External feedback may be used to suppress unexpected oscillations, e.g., to stabilize an array of coupled chaotic lasers to achieve a constant average output power [24]. It may also find application in ecology to stabilize ecological systems [25]. For its vanishing property, feedback induced oscillation death may be particularly suitable for these practical applications. This phenomenon may also arises in biological oscillator networks. An example is in pathologies of an assembly of cardiac pacemaker cells. Oscillation death means the cessation of regular cardiac rhythm. Previous studies ascribe such an arrhythmia either to a wide spread of natural frequencies of the cells [18, 19] or to a significant time delay in the signals exchanged between the cells [10]. Our work shows that neural control, which modulates the natural heart rate, may be responsible for such a situation [26]. Of course, only the simple all-to-all coupling strategy is considered here, feedback may lead to complex amplitude death in systems with complex network topology [27].

Figure Captions

Fig. 1  Evolution of oscillation death zone of the closed-loop system (2) in $(\Delta, \varepsilon)$-space with feedback gain $g$. $\alpha = 1.0$. (a) $g = 0$, (b) $g = 1.0$, (c) $g = 1.5$, and (d) $g = 5.0$.

Fig. 2  Bifurcation diagram of Equation (7) in $(\Delta, \varepsilon)$-space. (a) Without feedback, i.e. $g = 0$. The death zone (yellow region) is dependent on the distribution of $\omega_j$; (b) With strong feedback ($g > \alpha$). The parameter plane is divided by $\varepsilon = \alpha$ into death (yellow) and incoherence (red) zone.

Fig. 3  Two different ways to oscillation death. (a) Desynchronization induced oscillation death, and (b) Complete synchronization induced oscillation death in ensembles of identical oscillators.

Fig. 4  (a) Coupling induced synchronization in ensemble of chaotic Rossler oscillators ($N = 100$). Parameters $\omega_j$ are uniformly chosen in $[1-\gamma, 1+\gamma]$ with $\gamma = 0.01$. The coupling strength is $\varepsilon = 0.25$. (b) Death zone in $(\gamma, \varepsilon)$-space with strong feedback ($g = 2.0$). (c) Feedback induced oscillation death in an ensemble of identical coupled Rossler oscillators ($N = 100$), $\gamma = 0.0$, $g = 2.0$. (d) Time course of feedback signals $u(t)$. (e) Feedback induced oscillation death in an ensemble of non-identical coupled Rossler oscillators ($N = 100$), $\gamma = 0.15$, $g = 2.0$. (f) Time course of feedback signals $u(t)$.

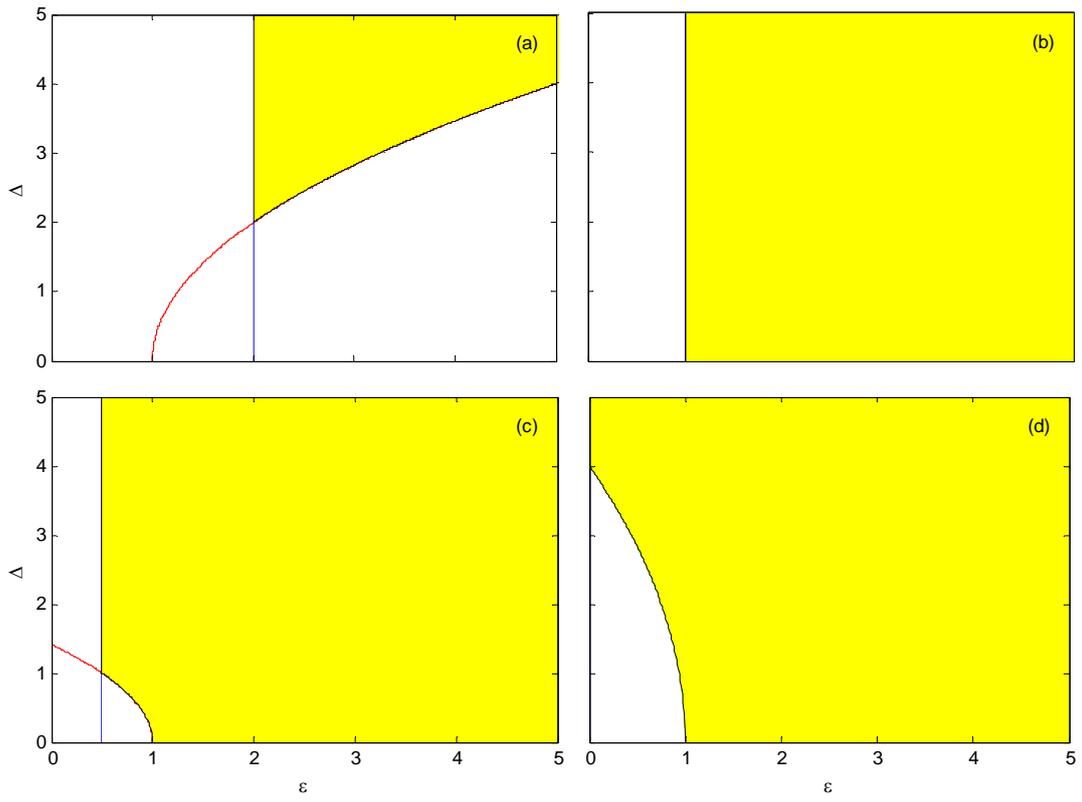

Fig. 1

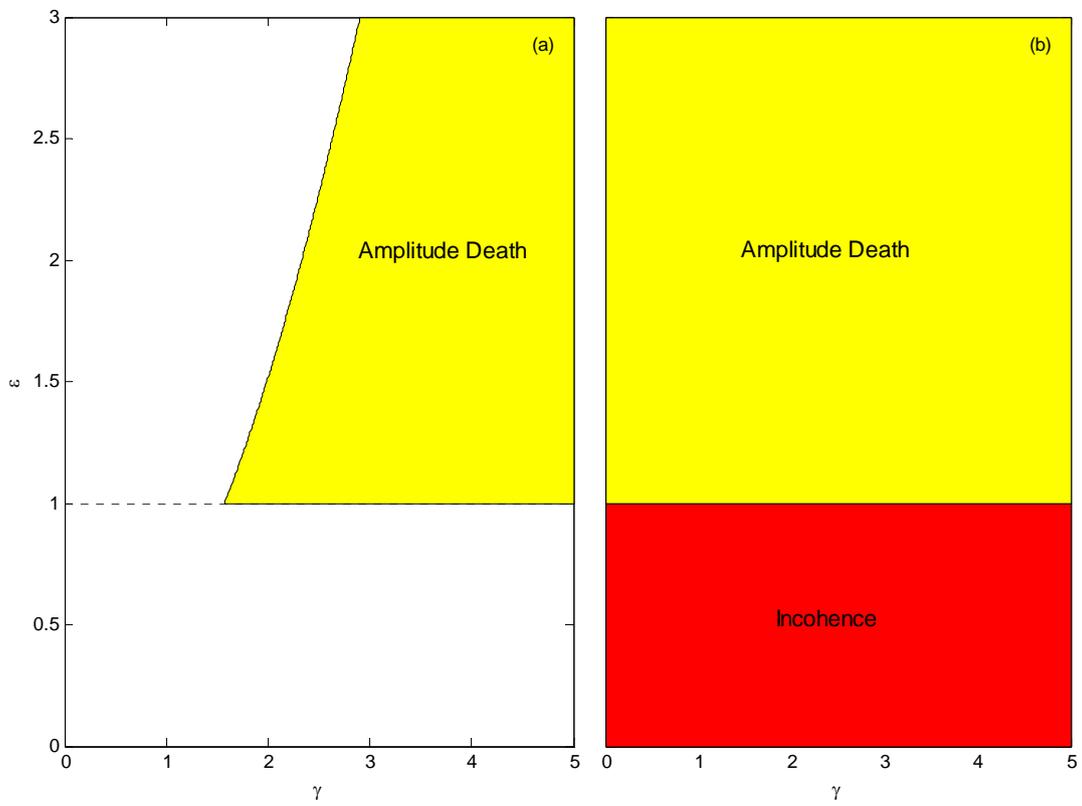

Fig. 2

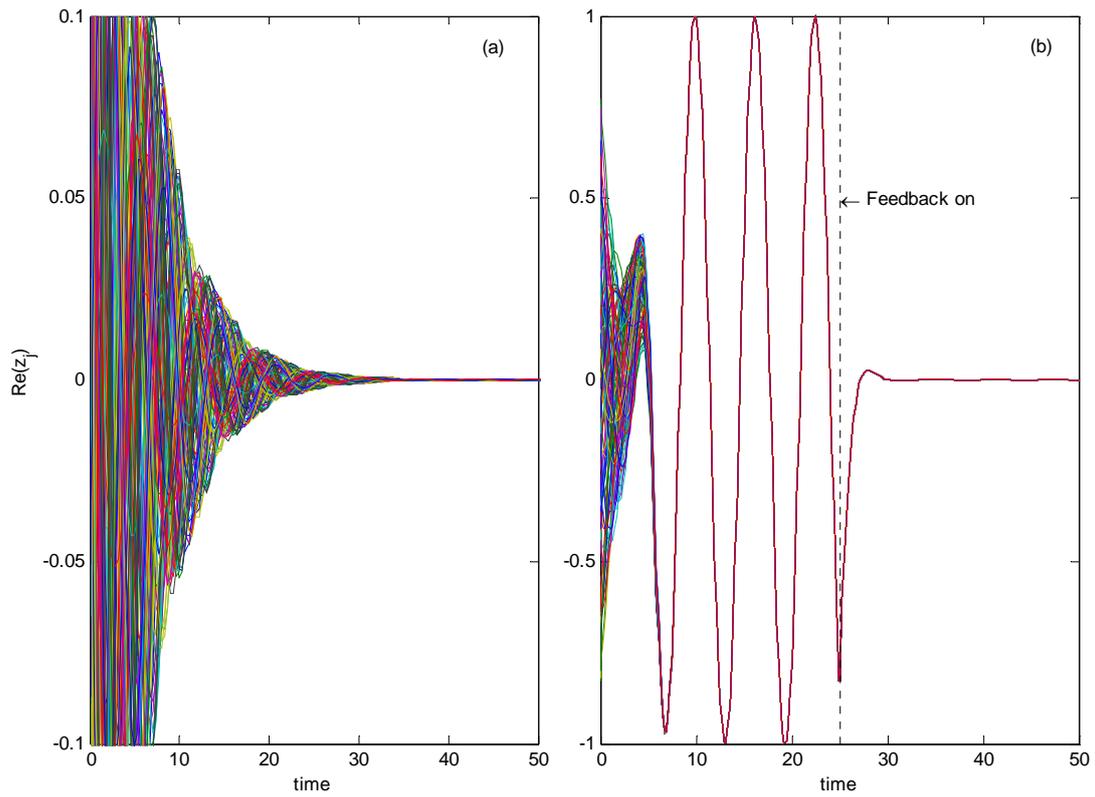

Fig. 3

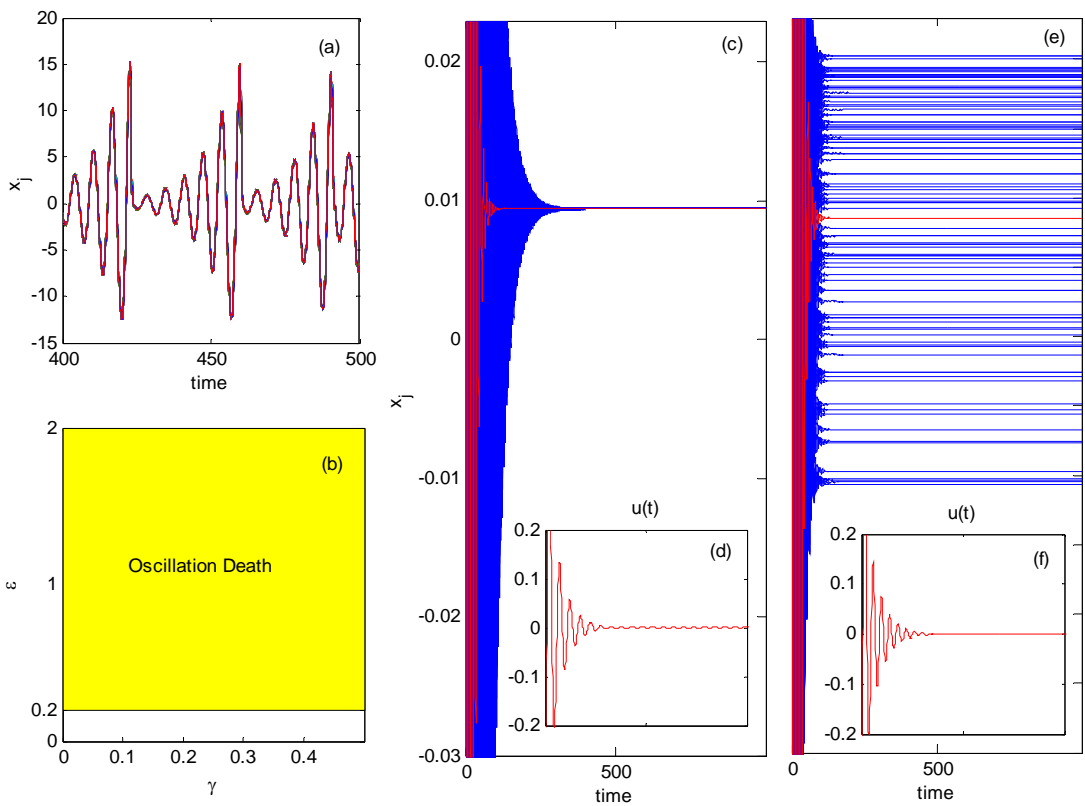

Fig. 4